\documentclass[12pt,notitlepage]{iopart}

\usepackage[noadjust]{cite}
\usepackage{graphicx,color}
\usepackage{amssymb}
\usepackage[colorlinks=true,linkcolor=blue,citecolor=blue]{hyperref}

\newcommand{\eqref}[1]{(\ref{#1})}

\newcommand{\cfrac}[2]{\displaystyle{\frac{#1}{#2}}}
\newcommand{\revfirst}[1]{\textcolor{black}{#1}}
\newcommand{\revsecond}[1]{\textcolor{black}{#1}}

\begin{document}

\title{Coherent dynamics in frustrated coupled parametric oscillators}

\author{Marcello~Calvanese~Strinati$^{1}$, Igal~Aharonovich$^2$, Shai~Ben-Ami$^2$, Emanuele~G.~Dalla~Torre$^{1}$, Leon~Bello$^{2}$, and Avi~Pe'er$^2$}
\address{$^1$Department of Physics, Bar-Ilan University, 52900 Ramat-Gan, Israel}
\address{$^2$Department of Physics and BINA Center of Nanotechnology, Bar-Ilan University, 52900 Ramat-Gan, Israel}


\begin{abstract}
We explore the coherent dynamics in a small network of three coupled parametric oscillators and demonstrate the effect of frustration on the persistent beating between them. Since a single-mode parametric oscillator represents an analog of a classical Ising spin, networks of coupled parametric oscillators are considered as simulators of Ising spin models, aiming to efficiently calculate the ground state of an Ising network - a computationally hard problem. However, the coherent dynamics of coupled parametric oscillators can be considerably richer than that of Ising spins, depending on the nature of the coupling between them (energy preserving or dissipative), as was recently shown for two coupled parametric oscillators. In particular, when the energy-preserving coupling is dominant, the system displays everlasting coherent beats, transcending the Ising description. Here, we extend these findings to three coupled parametric oscillators, focusing in particular on the effect of frustration of the dissipative coupling. We theoretically analyze the dynamics using coupled nonlinear Mathieu's equations, and corroborate our theoretical findings by a numerical simulation that closely mimics the dynamics of the system in an actual experiment. Our main finding is that frustration drastically modifies the dynamics. While in the absence of frustration the system is analogous to the two-oscillator case, frustration reverses the role of the coupling completely, and beats are found for small energy-preserving couplings. 
\end{abstract}


\section{Introduction}
Parametric oscillators are a viable experimental platform to study the physics of time crystals, i.e., systems that can spontaneously break time translational symmetry~\cite{sacha2017time,khemani019abriefhistory}. The possibility of the existence of such a phase of matter at equilibrium was first proposed in 2012 by Frank Wilczek and collaborators~\cite{PhysRevLett.109.160401,PhysRevLett.109.160402}, both for quantum and classical systems. The original proposal evokes the possibility for a system to break \emph{continuous} time translational symmetry, in analogy with the formation of space crystals in condensed matter where space translational symmetry is broken. Shortly after its proposal, it became clear that this kind of time-crystalline phase cannot exist at equilibrium~\cite{PhysRevLett.111.070402,Nozi_res_2013,PhysRevLett.114.251603}. However, following Wilczek's original idea, it was understood that time crystals can be realized out of equilibrium, in periodically-driven system, also referred to as Floquet systems. This new type of time crystals, dubbed \emph{Floquet time crystals}, accounts for the fact that, under certain conditions, a periodically-driven system can break the \emph{discrete} time translational symmetry enforced by the external drive~\cite{PhysRevA.91.033617,PhysRevLett.116.250401,PhysRevLett.117.090402,PhysRevB.94.085112,PhysRevB.96.115127,PhysRevLett.118.030401,PhysRevX.7.011026,yao2018classical,sullivan2018dissipative,doi:10.1063/PT.3.4020,PhysRevLett.122.015701}: Instead of merely following the external drive, the system undergoes a periodic motion at a frequency that is different from that of the drive (see ref.~\cite{sacha2017time} for a review).

The periodically driven single-mode classical parametric oscillator is the canonical example of period-doubling instability (see refs.~\cite{landau1982mechanics,strogatz2007nonlinear} for an introduction), and represents the simplest case of a classical Floquet time crystal. Indeed, when excited above the amplification threshold, the parametric oscillator oscillates at half the frequency of the drive and admits only \emph{two} distinct phase solutions, dubbed ``$0$'' and ``$\pi$'', with a relative shift in time by one period of the drive. One of the two solutions is chosen by the system depending on the initial conditions, a phenomenology analogous to a spontaneous breaking of a $\mathbb{Z}_2$ (Ising) symmetry. Because of this, a single degenerate parametric oscillator may be regarded as a classical bit, or an Ising spin, where the two states ``up'' or ``down'' of the spin are given by the two distinct ``$0$'' and ``$\pi$'' solutions. Exploiting this property, networks of many coupled parametric oscillators have been proposed as a platform, called \emph{coherent Ising machine} (CIM)~\cite{PhysRevA.88.063853}, to simulate the behaviour of a network of many coupled Ising spins. Such a machine, whose experimental realization has been reported in refs.~\cite{nphoton.2016.68,s41534-017-0048-9,takesue2018large2dising}, is envisioned to solve the NP-hard problem of finding the ground state of the classical Ising model~\cite{Barahona_1982}. \revfirst{In the last years, the analysis of various issues related to the computational performance of CIMs has been the focus of a remarkable amount of work, not only implementing CIMs using parametric-oscillator networks~\cite{Inagaki603,hamerlyfristratedchain2016,1805.05217,PhysRevLett.122.213902,PhysRevApplied.13.054059}, which are the focus of this paper, but also using digital computers~\cite{king2018emulating,Tiunov:19}, polariton networks~\cite{kalinin2019}, electrical oscillators~\cite{chou2019}, optoelectronical oscillators~\cite{bohm2019}, and laser networks~\cite{Utsunomiya2019}.}

While the underlying assumption in \revfirst{parametric-oscillator-based} CIMs \revsecond{(henceforth PO-CIMs)} is that a system of coupled parametric oscillators behaves as a set of coupled Ising spins, we pointed out recently that already a pair of coupled parametric oscillators may display a much richer dynamics, beyond the Ising description, depending on the nature of the coupling (energy-preserving or dissipative) between the oscillators~\cite{PhysRevLett.123.083901,PhysRevA.100.023835}. Specifically, we studied in detail both theoretically and experimentally in a radio-frequency experiment a pair of coupled parametric oscillators, which is the minimal system to explore nontrivial coupling effects. Our main finding was that, when driven above the amplification threshold at the parametric resonance condition, the two oscillators can either display \emph{persistent coherent beats} when the coupling is mostly energy preserving, or behave as a \revsecond{PO-CIM}~\cite{PhysRevA.88.063853} when the coupling is mostly dissipative.

The existence of such a nontrivial dynamics in just a pair of coupled parametric oscillators opens the question on how the nature of the coupling affects the dynamics of a larger network composed by more than two parametric oscillators, \revfirst{specifically with potential implications in the context of \revsecond{PO-CIM}s and} in the view of exploiting large-scale networks of coupled parametric oscillators to realize classical many-body time crystals~\cite{PhysRevLett.123.124301}. Motivated by these perspectives, we present in this paper a detailed theoretical and numerical analysis of three coupled parametric oscillators, which is the minimal system where nontrivial connectivity effects can be studied. The coupling between any two oscillators is parametrized by two coupling components - energy-preserving and dissipative. The main focus of this paper is to analyze \revfirst{for specific choices of the coupling matrix the effect of \emph{frustration} (defined as the situation where the dissipative couplings prevent the oscillators from adjusting their phases to energetically minimize every link~\cite{Vannimenus1977}), which turns out to be dramatic.}

To reach this goal, we model each parametric oscillator as a classical variable and describe the system by three coupled nonlinear Mathieu's equations~\cite{PhysRevA.100.023835}, in the presence of an external pump, intrinsic dissipation, and pump depletion nonlinearity, to analyze the phase diagram of the system for different values of the system parameters, as detailed hereon. Our theoretical predictions are confirmed by \revfirst{a} low-level numerical simulation of the field propagation within the parametric oscillators both in time and space, as close as possible to an actual experimental setup. Our numerical scheme emulates directly the dynamics of the field inside a cavity with parametric gain, with no explicit mention of the equations of motion studied in our analytical model.

Our main finding is that frustration totally inverts the dynamical picture of the coupled system. While in the absence of frustration the system behaves similar to the two-oscillator case, where beats are observed only when the energy-preserving coupling is larger than the dissipative one, in the presence of frustration we find two main differences: First, the system shows coherent everlasting beats for small energy-preserving couplings. This finding can be reasoned by the fact that a frustrated system cannot distinguish between two (or more) degenerate Ising states that are found when the coupling is purely dissipative. Thus, any non-vanishing value of the energy-preserving coupling induces beating between those degenerate states. Second, for large energy-preserving couplings and large frustration, the network converges to a phase-locked oscillation, which however is not the Ising ground state.

This paper is organized as follows. In section~\ref{sec:twoparametricoscillators}, we \revfirst{briefly} review \revfirst{our previous results of refs.~\cite{PhysRevLett.123.083901,PhysRevA.100.023835} for the simpler case of two coupled parametric oscillators}, introducing our model and notations. We then present our theoretical analysis for the case of three coupled parametric oscillators in section~\ref{sec:threecoupledparametricoscillators}. We discuss in section~\ref{sec:experimentalrealization} a possible experimental implementation of our system, and present the results of the low-level numerical simulation of such an experiment. We then draw our conclusions in section~\ref{sec:conclusions}, and report some relevant details on the calculations in the appendixes.


\section{Two parametric oscillators }
\label{sec:twoparametricoscillators}
Before moving to the analysis of three-coupled oscillators, let us \revfirst{shortly review} the relevant notation and analytical tools \revfirst{of our previous work in refs.~\cite{PhysRevLett.123.083901,PhysRevA.100.023835}. The familiar reader can directly move on to section~\ref{sec:threecoupledparametricoscillators} for the discussion on three-coupled parametric oscillators}.

\subsection{Model and notation}
\label{sec:modelandnotation}
\begin{figure}[t]
\centering
\includegraphics[width=8cm]{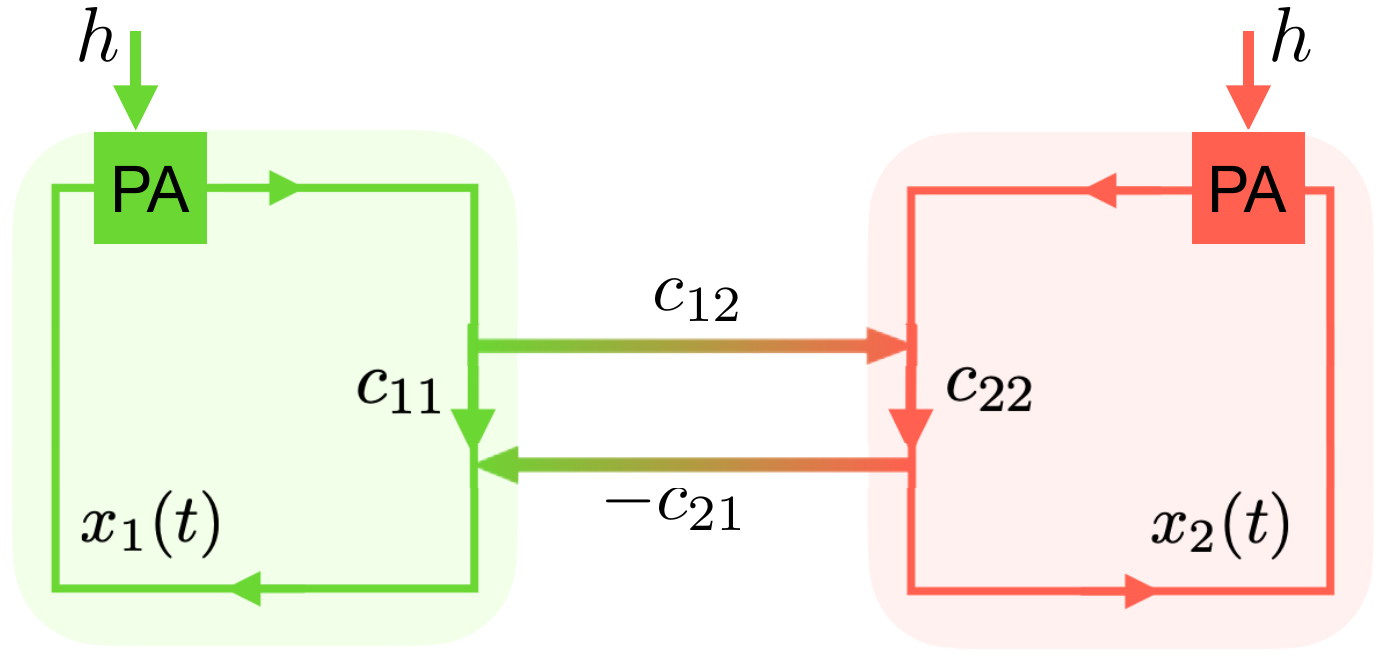}
\caption{Schematic representation of the two-parametric-oscillator system. Each parametric oscillator is described by a classical field $x_k$, with $k=1,2$, driven by an external pump $h(t)=h\,\sin(2\omega_0t)$ injected into a parametric amplifier (PA). The coupling between the oscillators is described in the general case by a coupling matrix that accounts for (i) transmittance coefficients $c_{11}$ and $c_{22}$, which renormalize the intrinsic losses of the oscillators, and (ii) coupling coefficients $c_{12}$ and $c_{21}$, which determine the rate of energy flow from oscillator 1 to 2, and \emph{vice versa}.}
\label{fig:schemetwooscillators}
\end{figure}

We consider a system of two identical single-mode parametric oscillators, with equal proper frequency $\omega_0$, driven by an external pump field at frequency $2\omega_0$ and with amplitude $h$, injected into a parametric amplifier (PA)~\cite{boyd2008nonlinear} as depicted in figure~\ref{fig:schemetwooscillators}. The field inside each oscillator 1 (or 2) is identified by a classical variable $x_1$ ($x_2$). The two oscillators are coupled by a power-splitter  coupling~\cite{PhysRevLett.123.083901}, which accounts for: (i) transmission coefficients $c_{11}$ and $c_{22}$ for oscillator 1 and 2, respectively, which renormalize the intrinsic loss of each oscillator, providing an overall loss rate that we denote by $g$, and (ii) coupling coefficients $c_{12}$ and $c_{21}$, which give the rate of energy exchange between the two oscillators. In this framework, the fields $x_1$ and $x_2$ are coupled according to the equation (see \ref{appendix:derivationofthepowersplittercoupling})
\begin{equation}
\left(\begin{array}{c}\dot x_1\\\dot x_2\end{array}\right)=
\omega_0\left(\begin{array}{cc}0&c_{12}\\-c_{21}&0\end{array}\right)
\left(\begin{array}{c}x_1\\ x_2\end{array}\right) \,\, ,
\label{eq:powersplittercouplingmatrix}
\end{equation}
where the dot denotes the time derivative. The dynamics of the two-oscillator system is described by a pair of coupled Mathieu's equations~\cite{PhysRevA.100.023835}
\begin{equation}
\left\{\begin{array}{l}
\ddot x_1+\omega_0^2\left[1+h(1-\beta\,x_1^2)\sin(2\omega_0t)\right]\,x_1+\omega_0\,g\,\dot x_1-\omega_0\,c_{12}\,\dot x_2=0\\\\
\ddot x_2+\omega_0^2\left[1+h(1-\beta\,x_2^2)\sin(2\omega_0t)\right]\,x_2+\omega_0\,g\,\dot x_2+\omega_0\,c_{21}\,\dot x_1=0
\end{array}\right. \,\, .
\label{eq:equationsofmotionfortwooscillators}
\end{equation}
Equation~\eqref{eq:equationsofmotionfortwooscillators} also includes a second-order nonlinearity in the amplitude of the pump field (hereafter referred to as ``pump-depletion nonlinearity''), whose strength is quantified by $\beta$. Such a nonlinearity describes the fact that the intensity of the pump field inside each oscillator is depleted by $x_1$ and $x_2$, and in many experimental contexts captures the most relevant nonlinear process~\cite{PhysRevA.100.023835}.
\revfirst{In general}, the rate of energy flow between the two oscillators can be unbalanced, i.e., $c_{12}\neq c_{21}$, indicating dissipation in the coupling itself. Without loss of generality, one can parametrize the coupling coefficients \revfirst{in terms of an antisymmetric and symmetric part with respect to the exchange $x_1\leftrightarrow x_2$ in Eq.~\eqref{eq:equationsofmotionfortwooscillators}}: $c_{12}=r-\alpha$ and $c_{21}=r+\alpha$, where \revfirst{the antisymmetric part} $r\geq0$ represents the \emph{energy-preserving} component of the coupling, whereas \revfirst{the symmetric part} $\alpha\geq0$ is the \emph{dissipative} one.

The energy-preserving coupling $r$ induces a coherent exchange of energy between the two oscillators. Its energy-preserving nature follows from the fact that the equations of motion~\eqref{eq:equationsofmotionfortwooscillators}, with $\beta=0$, $g=0$, and $c_{12}=c_{21}=r$, can be derived from the Hamilton's equations~\cite{landau1982mechanics} starting from the Hamiltonian
\begin{equation}
\hspace{-1.5cm}H=\frac{p^2_1+p^2_2}{2m}+\frac{1}{2}\,m\,\omega^2_0\left[1+\frac{r^2}{4}+h\,\sin(2\omega_0t)\right]\left(x_1^2+x_2^2\right)+\frac{\omega_0r}{2}\left(p_1x_2-p_2x_1\right) \,\, ,
\label{eq:hamiltonianenergypreservingcoupling}
\end{equation}
where $p_1$ and $p_2$ are the canonical momentum variables for $x_1$ and $x_2$, respectively. Such an Hamiltonian is analogous to that of a charged particle (charge $q=m\omega_0r/2$) moving on a two-dimensional plane identified by the spatial coordinates $(x_1,x_2,z=0)$, subject to a vector potential $\mathbf{A}={(-x_2,x_1,0)}^T$, where $T$ denotes the transposition (for the details of the derivation, see~\ref{appendix:hamiltonianforthepowersplittercoupling}).

The dissipative coupling $\alpha$, in contrast, introduces additional loss or gain terms~\cite{PhysRevA.100.023835}, which give rise to the Ising-type coupling between the oscillators \revfirst{that is usually considered in the standard analysis of \revsecond{PO-CIM}s~\cite{PhysRevA.88.063853,nphoton.2016.68,s41534-017-0048-9,takesue2018large2dising,Inagaki603,hamerlyfristratedchain2016,1805.05217,PhysRevLett.122.213902,PhysRevApplied.13.054059}}. \revfirst{This coupling} guides the convergence of the two-oscillator system to the desired Ising ground state~\cite{PhysRevA.100.023835}. In the long-time limit, the two oscillators will prefer to lock according to the sign of $\alpha$: In-phase for ``ferromagnetic'' coupling ($\alpha>0$), yielding the two ``ferromagnetic'' configurations $(00)$ or $(\pi\pi)$, or in anti-phase for ``anti-ferromagnetic'' coupling ($\alpha<0$), yielding the two ``anti-ferromagnetic'' configurations $(0\pi)$ or $(\pi0)$, where the first (second) label denotes the corresponding phase solution the first (second) oscillator.

\subsection{Phase-locking and beats}
We now review the effect of the interplay between $r$ and $\alpha$ on the long-time dynamics of the system. \revfirst{We focus on the dynamics of the slow-varying amplitudes that modulate the fast-varying oscillations at half the pump frequency, which are instead integrated out, by employing the multiple-scale perturbative expansion in~\cite{PhysRevA.100.023835}}. \revfirst{We take the intra-cavity loss $g$ as a small expansion parameter, and identify the fast-varying and slow-varying time scales as $t=2\pi/\omega_0$ and $\tau=gt$, respectively. By writing $x_1(t,\tau)=A(\tau)\,e^{i\omega_0t}+A^*(\tau)\,e^{-i\omega_0t}$ and $x_2(t,\tau)=B(\tau)\,e^{i\omega_0t}+B^*(\tau)\,e^{-i\omega_0t}$, where $A(\tau)$ and $B(\tau)$ are the complex amplitudes that encode the slow-varying dynamics, and} by \revfirst{rescaling} \revfirst{$\tilde h=h/g$, $\tilde r=r/g$, $\tilde \alpha=\alpha/g$, and} $\tilde\tau=\omega_0\tau$, one finds that $A$ and $B$ obey the following set of coupled first-order differential equations~\cite{PhysRevA.100.023835}:
\begin{eqnarray}
\frac{\partial A}{\partial\tilde\tau}&=&\frac{\tilde h}{4}\,A^*-\frac{\tilde h\,\beta}{4}\left(3{|A|}^2A^*-A^3\right)-\frac{A}{2}+\frac{\tilde r+\tilde\alpha}{2}\,B\nonumber\\\nonumber\\
\frac{\partial B}{\partial\tilde\tau}&=&\frac{\tilde h}{4}\,B^*-\frac{\tilde h\,\beta}{4}\left(3{|B|}^2B^*-B^3\right)-\frac{B}{2}-\frac{\tilde r-\tilde\alpha}{2}\,A \,\, ,
\label{eq:nonlinearmathieuequationflowcoupled}
\end{eqnarray}
Equation~\eqref{eq:nonlinearmathieuequationflowcoupled} can be further recast in terms of the real and imaginary parts of the complex amplitudes, $A=A_R+i\,A_I$ and $B=B_R+i\,B_I$, where $A_R$ ($B_R$) and $A_I$ ($B_I$) are, respectively, the real and imaginary parts of $A$ ($B$). \revfirst{The long-time dynamics is determined by the configuration and stability of the fixed points $(\overline{A}_R,\overline{A}_I,\overline{B}_R,\overline{B}_I)$ of equation~\eqref{eq:nonlinearmathieuequationflowcoupled}. Note however that, sufficiently close to the oscillation threshold, $A_I$ and $B_I$ decay very quickly ($\overline{A}_I=\overline{B}_I=0$)~\cite{PhysRevA.100.023835} due to the phase dependent amplification and squeezing in parametric oscillators, allowing to focus the discussion only on the dynamics of the real parts $A_R$ and $B_R$}.

While in general the configuration of the fixed points depends on the form of the nonlinearity, especially far from the amplification threshold, most of the interesting physics throughout this paper occurs close to the threshold, where nonlinear effects are negligible and the system is almost linear.  The properties of the system at threshold can be therefore found by \revfirst{focusing on the spectrum of the Jacobian matrix around the origin $A=B=0$, analyzing specifically the eigenvalue with largest real part ($\lambda_{\rm max}$, which we dub ``most efficient eigenvalue'' from now on).} \revfirst{Importantly, when the system exceeds the amplification threshold $\tilde h_{\rm th}$, defined by the condition ${\rm Re}[\lambda_{\rm max}]=0^+$, the imaginary part of $\lambda_{\rm max}$ determines the \emph{frequency of the beats} at threshold $\omega_{\rm B}=\omega_0g|{\rm Im}[\lambda_{\rm max}]|$ between the two oscillators}. The beat frequency $\omega_{\rm B}$ is the key observable to describe the behaviour of the system as the oscillators are driven above the oscillation threshold. \revfirst{Specifically}, when $\omega_{\rm B}=0$, $A$ and $B$ reach eventually constant values $\overline A_R$ and $\overline B_R$, and the system behaves as a time crystal, and can simulate Ising spins. Indeed, \revfirst{for positive (negative) $\overline{A}_R$}, $x_1$ converges to the ``$0$'' \revfirst{(``$\pi$'')} solution, and analogously for $x_2$. Instead, for $\omega_{\rm B}>0$, $A$ and $B$ display \emph{persistent coherent beats}. The presence of the beats implies that each oscillator $x_{1,2}$ periodically flips between the ``$0$'' and ``$\pi$'' solutions, and therefore the system neither obeys the Ising description, nor it behaves as a time crystal.

\begin{figure}[t]
\centering
\includegraphics[width=14cm]{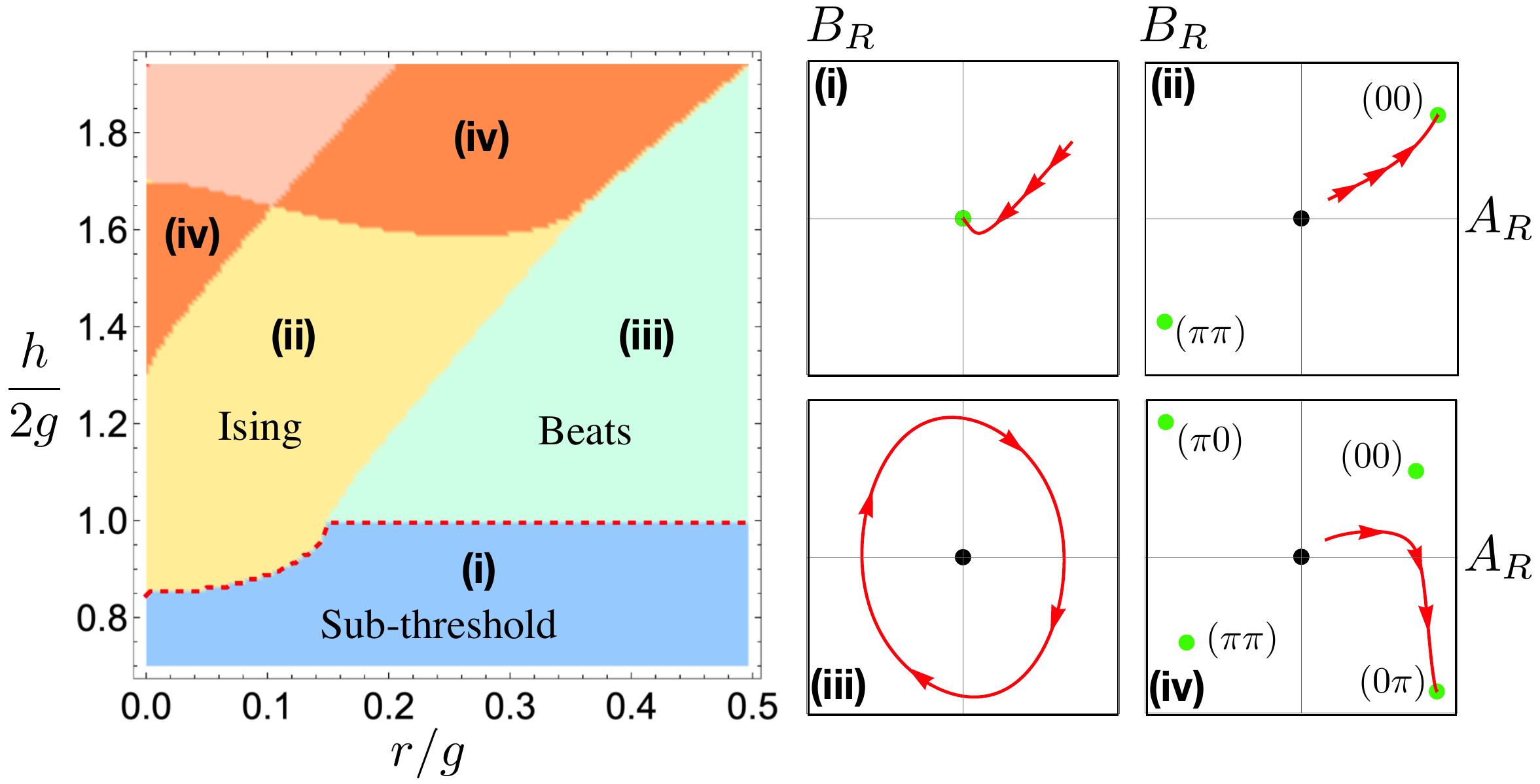}
\caption{Dynamical phase diagram of two coupled parametric oscillators, described by equation~\eqref{eq:nonlinearmathieuequationflowcoupled}. (Left) Phase diagram in the $h/(2g)$ vs. $r/g$ plane, for $\tilde\alpha=0.15$, and (Right) configuration of the fixed points in the $B_R$ vs. $A_R$ plane \revfirst{(while $A_I=B_I=0$)}, where black and green dots represent unstable and stable points, respectively. The flow of the slow varying amplitudes from the solution of equation~\eqref{eq:nonlinearmathieuequationflowcoupled} is marked by red curves. The phase diagram consists of four main phases, characterized by different configurations of the fixed points, as shown in right panels: \textbf{(i)} below the oscillation threshold, where only the origin $A=B=0$ is a stable attractor; \textbf{(ii)} the Ising or \revsecond{PO-CIM} region, where the system has two stable fixed points, corresponding to the two ground-state Ising solutions $(00)$ and $(\pi\pi)$ (for ``ferromagnetic'' $\tilde\alpha>0$); \textbf{(iii)} a phase in which a stable limit cycle stabilizes the long-time dynamics, and the system displays coherent beats; \textbf{(iv)} a phase with four stable fixed points, corresponding to both ground-state $(00)$ and $(\pi\pi)$, and excited-state $(0\pi)$ and $(\pi0)$ configurations. The red dashed line in the phase diagram is the boundary for the oscillation threshold $\tilde h_{\rm th}$. Other phases with more than four fixed points~\cite{PhysRevA.100.023835} are not labelled and not relevant for the present discussion, and additional unstable fixed points different from the origin are not shown for the sake of clarity.}
\label{fig:phasediagramtwooscillators}
\end{figure}

\revfirst{A} concrete calculation of the phase diagram of the system in equation~\eqref{eq:nonlinearmathieuequationflowcoupled} \revfirst{is shown in figure~\ref{fig:phasediagramtwooscillators}, where we identify the following main phases:}
\begin{itemize}
\item[\textbf{(i)}] The sub-threshold phase, for a pump amplitude $\tilde h < \tilde h_{\rm th}$, where the origin $A=B=0$ is the only stable fixed point of equation~\eqref{eq:nonlinearmathieuequationflowcoupled};
\item[\textbf{(ii)}] The Ising or \revsecond{PO-CIM} phase, for $\tilde h>\tilde h_{\rm th}$, \revfirst{where} \emph{two} stable fixed points \revfirst{are found}, the origin being unstable. \revfirst{Phase locking occurs} at $(00)$ or $(\pi\pi)$, for ``ferromagnetic'' $\tilde\alpha>0$, or at $(0\pi)$ or $(\pi0)$ for ``antiferromagnetic'' $\tilde\alpha<0$ (not shown);
\item[\textbf{(iii)}] The beating phase, for $\tilde h>\tilde h_{\rm th}$, where the long-time dynamics \revfirst{of the oscillators' amplitudes} is attracted into a \emph{stable limit cycle};
\item[\textbf{(iv)}] A phase with four stable fixed points, corresponding to all the four Ising configurations $(00)$, $(\pi\pi)$, $(0\pi)$, and $(\pi0)$, in which the system behaves as two \emph{decoupled} spins. This behaviour matches the one previously discussed in ref.~\cite{PhysRevA.88.063853}.
\end{itemize}
Slightly above the amplification threshold $\tilde h_{\rm th}$, the \revsecond{PO-CIM} phase is found for $\tilde r<\tilde\alpha$ ($\omega_{\rm B}=0$), whereas the beating phase is found for $\tilde r>\tilde\alpha$ ($\omega_{\rm B}>0$). At threshold, the system therefore undergoes a transition between the \revsecond{PO-CIM} and the beating behaviour when $\tilde r=\tilde\alpha$.


\begin{figure}[t]
\centering
\includegraphics[width=13cm]{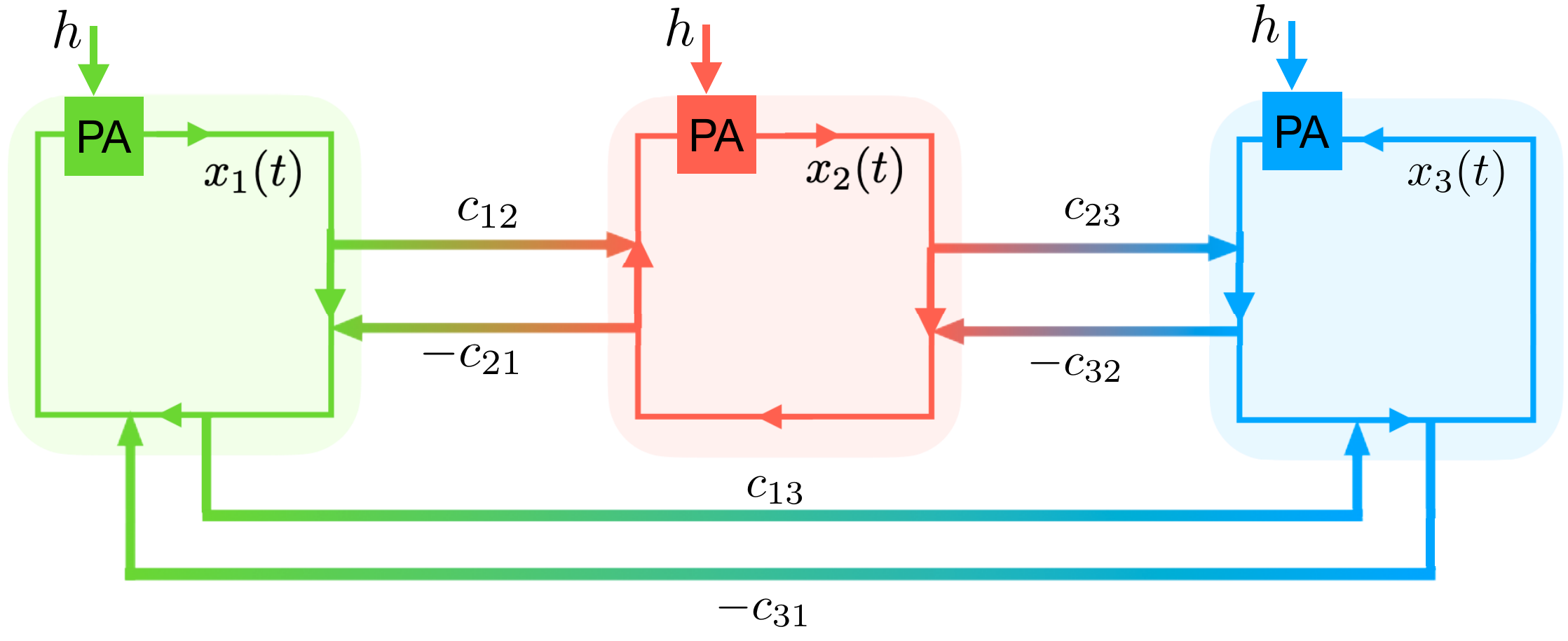}
\caption{Schematic representation of the three-parametric-oscillator system. Similar to figure~\ref{fig:schemetwooscillators}, the parametric oscillators are described by a classical field $x_k$ ($k=1,2,3$), driven by the same external pump $h(t)=h\,\sin(2\omega_0t)$. The coupling between the oscillators, describing the connectivity of the system, includes all the mutual couplings $c_{jk}$ for $j,k=1,2,3$. The transmittance coefficients $c_{jj}$ for $j=1,2,3$ have been omitted from the figure for the sake of simplicity.}
\label{fig:schemethreeoscillators}
\end{figure}

\section{Three coupled parametric oscillators}
\label{sec:threecoupledparametricoscillators}
The results \revfirst{of refs.~\cite{PhysRevLett.123.083901,PhysRevA.100.023835}, reviewed} in the previous section\revfirst{,} pointed out the existence of a persistent coherent beating dynamics in coupled parametric oscillators, not considered in the standard analysis of \revsecond{PO-CIM}s. A natural question that arises is how the presence of such a dynamics affects a more structured network with more than two coupled parametric oscillators. Here, we begin to address this question by extending the previous discussion to the case of three degenerate coupled parametric oscillators, which is the simplest configuration where one can systematically study the role of connectivity. Specifically, our main focus is to study how frustration of the dissipative coupling (which reflects the underlying Ising model) affects the coherent dynamics of the system. 

\subsection{Model}
\label{sec:threeparametricoscillatormodel}
A schematic representation of a three-oscillator system is shown in figure~\ref{fig:schemethreeoscillators}. Now, the coupling matrix $\mathbf{c}$ between the oscillators, according to equation~\eqref{eq:powersplittercouplingmatrix}, is
\begin{equation}
\left(
\begin{array}{c}\dot x_1\\\dot x_2\\\dot x_3\end{array}\right)=\omega_0
\left(\begin{array}{ccc}
0 & c_{12} & c_{13}\\
-c_{21} & 0 & c_{23}\\
-c_{31} & -c_{32} & 0
\end{array}\right)
\left(\begin{array}{c}x_1\\x_2\\x_3\end{array}\right) \,\, .
\label{eq:couplingmatrixthreecoupledparametricoscillators}
\end{equation}
The system is now described by a set of three coupled nonlinear Mathieu's equations, which we write in a compact form for the sake of simplicity ($j,k=1,2,3$):
\begin{equation}
\hspace{-2.2cm}
\ddot x_j+\omega_0^2\left[1\!+\!h\left(1\!-\!\beta\,x_j^2\right)\!\sin(2\omega_0t)\right]x_j+\omega_0g\dot x_j-\omega_0\!\sum_{k\neq j}{\rm sgn}(k\!-\!j)c_{jk}\dot x_k=0  \,\, ,
\label{eq:threecoupledparametricoscillatorequationsofmotionglobal}
\end{equation}
where ${\rm sgn}(\cdot)$ denotes the sign function. From equation~\eqref{eq:threecoupledparametricoscillatorequationsofmotionglobal}, one obtains the corresponding multiple-scale equations for the slow-varying amplitudes of the fields $\{x_j\}$ as in \revfirst{equation~\eqref{eq:nonlinearmathieuequationflowcoupled}}. Here, we renormalize each element of the coupling matrix as $\tilde c_{jk}=c_{jk}/g$, and to ease the notation, we use the symbols $\{X_j(\tau)\}$ of the amplitudes, such that $x_j(t,\tau)=X_j(\tau)e^{i\omega_0 t}+X^*_j(\tau)e^{-i\omega_0 t}$. The equations for the complex amplitudes are then determined ($j,k=1,2,3$):
\begin{equation}
\hspace{-1.2cm}\frac{\partial X_j}{\partial\tilde\tau}=\frac{\tilde h}{4}X_j^*-\frac{1}{2}X_j-\frac{\tilde h\beta}{4}\left(3{|X_j|}^2X_j^*-X_j^3\right)+\frac{1}{2}\sum_{k\neq j}{\rm sgn}(k-j)\tilde c_{jk}X_k\,\,.
\label{eq:equationsforamplitudesthreeparametricoscillator}
\end{equation}
As in section~\ref{sec:twoparametricoscillators}, we decompose the coupling $c_{jk}$ in equations~\eqref{eq:couplingmatrixthreecoupledparametricoscillators}-\eqref{eq:equationsforamplitudesthreeparametricoscillator} \revfirst{between any two oscillators} in terms of an \revfirst{energy-preserving (antisymmetric)} and \revfirst{dissipative (symmetric)} part, respectively $r_{jk}$ and $\alpha_{jk}$. Due to this increase of parameter space with respect to the case in section~\ref{sec:twoparametricoscillators} (the coupling matrix now has in general six independent components), we focus on a specific choice of the coupling matrix, with the ambition to highlight the role of frustration in the dissipative components of the coupling matrix. We choose $r_{jk}=r$ for all $j$ and $k$, and introduce two different dissipative couplings: $\alpha_{12}=\eta$ and $\alpha_{13}=\alpha_{23}=\alpha$, so that the coupling matrix in equation~\eqref{eq:couplingmatrixthreecoupledparametricoscillators} reads
\begin{equation}
\left(
\begin{array}{c}\dot x_1\\\dot x_2\\\dot x_3\end{array}\right)=
\omega_0g\left(\begin{array}{ccc}
0 & \tilde r+\tilde\eta & \tilde r+\tilde\alpha\\
-\tilde r+\tilde\eta& 0 & \tilde r+\tilde\alpha\\
-\tilde r+\tilde\alpha & -\tilde r+\tilde\alpha & 0
\end{array}\right)
\left(\begin{array}{c}x_1\\x_2\\x_3\end{array}\right) \,\, .
\label{eq:couplingmatrixthreecoupledparametricoscillators2}
\end{equation}
In the rest of the paper, our goal is to study the physics of the system near threshold as a function of the coupling parameters $\tilde r$, $\tilde\alpha$, and $\tilde\eta$, where as before all quadratures are real ($\overline{X}_{1,I}=\overline{X}_{2,I}=\overline{X}_{3,I}=0$). Before discussing this general case, we first focus on two main configurations of interest for the coupling matrix in equation~\eqref{eq:couplingmatrixthreecoupledparametricoscillators2}. Namely, assuming $\tilde r,\tilde\alpha\geq0$: (i) the \emph{non-frustrated} case, for $\tilde\eta=\tilde\alpha$, and (ii) the \emph{fully-frustrated} case, for $\tilde\eta=-\tilde\alpha$. The reason why we focus on these two fine-tuned cases is because they are, on one hand, easily analytically tractable, and on the other hand, they capture the dramatic effect of frustration in the dissipative coupling. We discuss the general case later in section~\ref{sec:generalcase}.

\subsection{Non-frustrated network}
\label{sec:nonfrustratednetwork}
First, we analytically discuss the threshold properties of the non-frustrated network, for $\tilde\eta=\tilde\alpha$ in equation~\eqref{eq:couplingmatrixthreecoupledparametricoscillators2}. We analyze separately the cases of $\tilde r>\tilde\alpha$ and $\tilde r<\tilde\alpha$.

For $\tilde r>\tilde\alpha$, we have $\lambda_{\rm max}=-1/2+\tilde h/4-e^{i\\\pi/3}\,F(\tilde r,\tilde\alpha)+e^{-i\,\pi/3}\,G(\tilde r,\tilde\alpha)$, where
\begin{equation}
F(\tilde r,\tilde\alpha)=\cfrac{\tilde r^2-\tilde\alpha^2}{4\,G(\tilde r,\tilde\alpha)} \qquad G(\tilde r,\tilde\alpha)=\cfrac{1}{2}\,{\left[\left(\tilde r^2-\tilde\alpha^2\right)\left(\tilde r+\tilde\alpha\right)\right]}^{1/3} \,\, .
\label{eq:expressionsforfandg}
\end{equation}
Since $\lambda_{\rm max}$ is complex, beats are found. From the imaginary parts of $\lambda_{\rm max}$, one can find the expression of the beat frequency:
\begin{equation}
\omega_{B}=\frac{g\omega_0\sqrt{3}}{4}{\left(\tilde r^2-\tilde\alpha^2\right)}^{1/3}\left[{\left(\tilde r-\tilde\alpha\right)}^{1/3}+{\left(\tilde r+\tilde\alpha\right)}^{1/3}\right] \label{eq:frequencyofthebeatsnonfrustrated}\,\, .
\end{equation}
In particular, for $\tilde r$ approaching $\tilde\alpha$, the frequency of the beats reduces towards zero with the critical behaviour of $\omega_B\sim{(\tilde r-\tilde\alpha)}^{1/3}$, differently from the critical exponent of $1/2$ in the two-oscillator case~\cite{PhysRevA.100.023835}.

\begin{figure}[t]
\centering
\includegraphics[width=15cm]{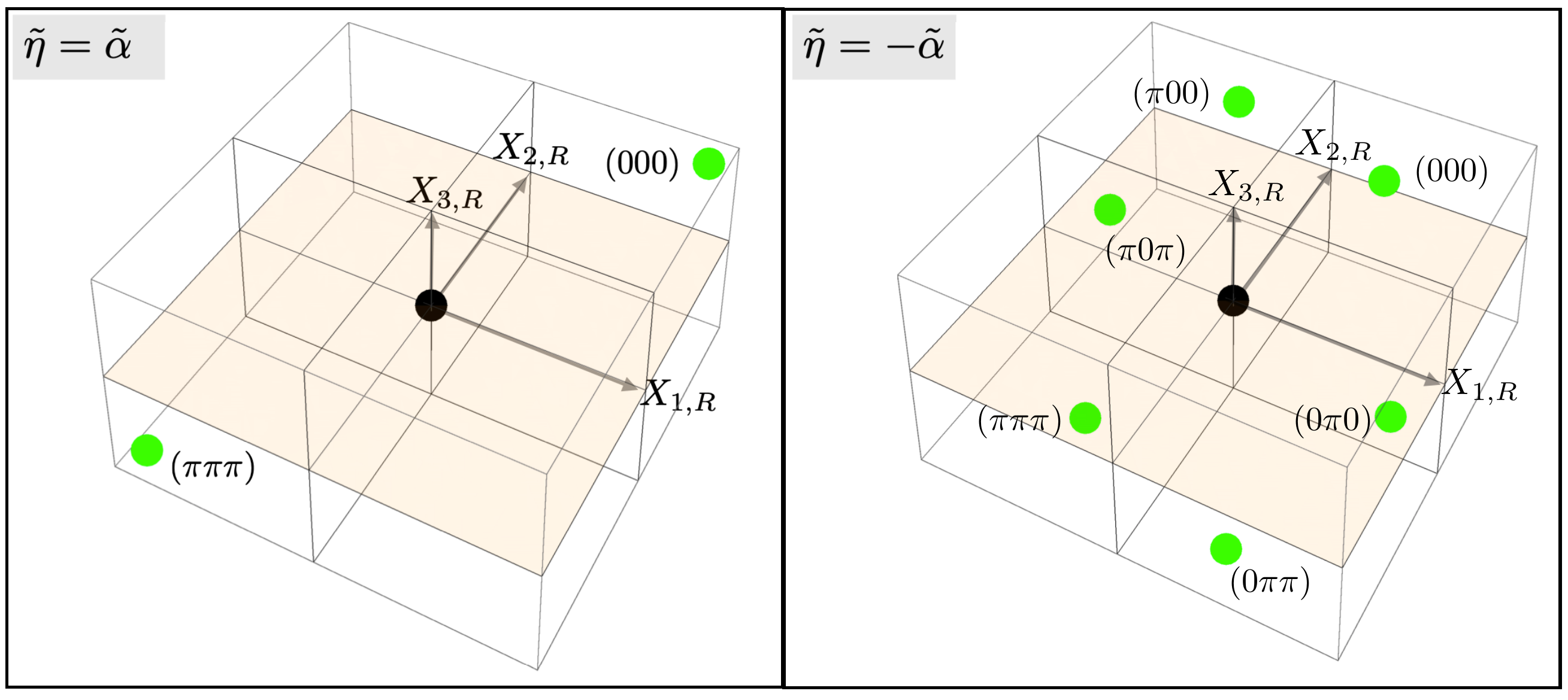}
\caption{Stability diagram of three parametric oscillators in the $(X_{1,R},X_{2,R},X_{3,R})$ space. Green dots represent stable configurations of oscillations for $\tilde r=0$ and $\tilde\alpha>0$. The unstable origin is represented by a black dot. The orange plane marks the $X_{3,R}=0$ plane. (Left panel) In the non-frustrated network ($\tilde\eta=\tilde\alpha$), the two possible phase-locked configurations are $(000)$ and $(\pi\pi\pi)$, which correspond to the ferromagnetic Ising solutions. The system converges to one of them as long as $\tilde r<\tilde\alpha$ ($\omega_{\rm B}=0$ for $\tilde r<\tilde\alpha$). (Right panel) In the frustrated case ($\tilde\eta=-\tilde\alpha$) there are six possible phase-locked configurations: $(0\pi\pi)$, $(0\pi0)$, $(000)$, $(\pi00)$, $(\pi0\pi)$, and $(\pi\pi\pi)$, which correspond to the six degenerate Ising ground states. In contrast to the non-frustrated case, any infinitesimal $\tilde r>0$ induces beats within these configurations ($\omega_{\rm B}>0$ for $0<\tilde r<\tilde\alpha$).}
\label{fig:fixedpointconfigurationsthreeoscillators}
\end{figure}

When $\tilde r<\tilde\alpha$, we find that $\lambda_{\rm max}=-1/2+\tilde h/4+F(\tilde \alpha,-\tilde r)+G(\tilde \alpha,-\tilde r)$. Now, $\lambda_{\rm max}$ is real, and above the oscillation threshold, parametric amplification occurs without beats, $\omega_B=0$. In this case, the phase-locked steady-state oscillations correspond to the two ``ferromagnetic'' configurations $(000)$ or $(\pi\pi\pi)$, as shown in the left panel figure~\ref{fig:fixedpointconfigurationsthreeoscillators} (in the figure specifically for $\tilde r=0$). As one may expect, the behaviour of the non-frustrated network is qualitatively the same as the behaviour of two coupled oscillators (section~\ref{sec:twoparametricoscillators}).

\subsection{Fully-frustrated network}
\label{sec:fullyfrustratednetwork}
We now move to the case of the fully-frustrated network, i.e., $\tilde\eta=-\tilde\alpha$ in equation~\eqref{eq:couplingmatrixthreecoupledparametricoscillators2}.  By proceeding as before, we find that, for $\tilde r>\tilde\alpha$, the most efficient eigenvalue is $\lambda_{\rm max}=-1/2+\tilde h/4+F(\tilde r,-\tilde\alpha)-G(\tilde r,-\tilde\alpha)$. Now, in stark contrast to the non-frustrated case, $\lambda_{\rm max}$ is real, and above the oscillation threshold parametric amplification occurs without beats, $\omega_B=0$.

\begin{figure}[t]
\centering
\includegraphics[width=15cm]{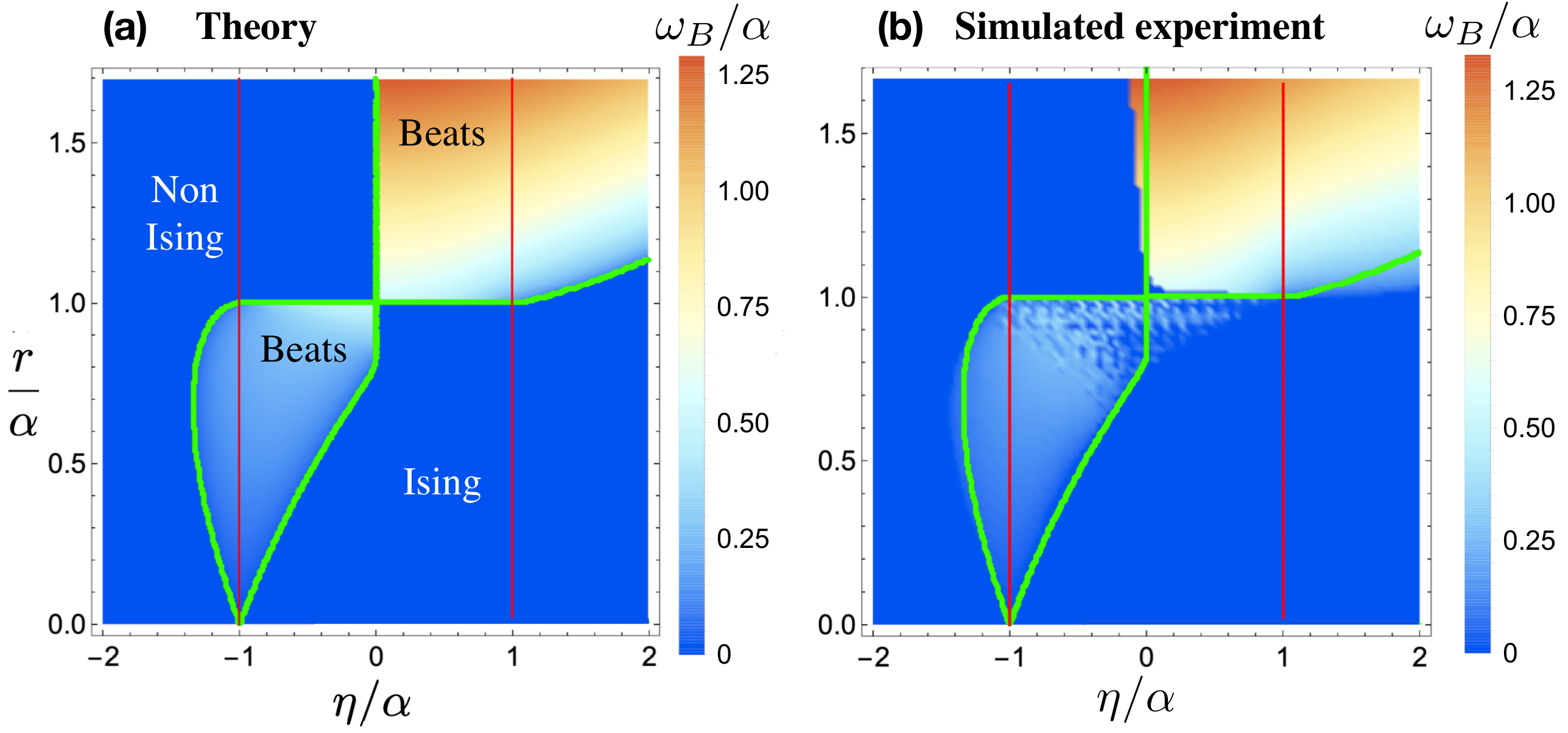}
\caption{Phase diagram of the three-oscillator case, showing the colormap plots of the frequency of the beats $\omega_B/\alpha$ at threshold in the $r/\alpha$ vs. $\eta/\alpha$ plane. The vertical red lines mark the values $\eta=\pm\alpha$ of non-frustrated and fully-frustrated network (see sections~\ref{sec:nonfrustratednetwork} and~\ref{sec:fullyfrustratednetwork}), and green lines are calculated boundaries between the phase-locking regions where $\omega_B=0$ (dark blue) and the beating regions where $\omega_B>0$ (other colors, see legends). \textbf{(a)} Analytical value from the theory presented in sections~\ref{sec:generalcase} and~\ref{appendix:jacobianmatrixspectrumintheinterpolatingcase}. The two phase-locking regions are labelled by ``Ising'' and ``Non Ising'' depending on whether the systems behaves correctly as a \revsecond{PO-CIM} or not (see section~\ref{sec:generalcase}). \textbf{(b)} Low-level simulation of the experiment (see section~\ref{sec:experimentalrealization}). The experimental simulation agrees exceptionally well with the predicted theoretical behaviour, apart from small deviations (some noisy regions and small discrepancies near the phase boundaries) that we ascribe mostly to the difficulty of estimating the oscillation threshold in the experimental simulation for some values of the system parameters.}
\label{fig:frequencybeatsrvseta}
\end{figure}

For $\tilde r<\tilde\alpha$, instead, $\lambda_{\rm max}$ reads as $\lambda_{\rm max}=-1/2+\tilde h/4+e^{i\,\pi/3}\,F(\tilde\alpha,\tilde r)+e^{-i\,\pi/3}\,G(\tilde\alpha,\tilde r)$. Therefore, the parametric oscillation occurs with beats, where the beat frequency is
\begin{equation}
\omega_{B}=\frac{g\omega_0\sqrt{3}}{4}{\left(\tilde\alpha^2-\tilde r^2\right)}^{1/3}\left[-{\left(\tilde\alpha-\tilde r\right)}^{1/3}+{\left(\tilde\alpha+\tilde r\right)}^{1/3}\right] \label{eq:frequencyofthebeatsfrustrated}\,\, .
\end{equation}
One can see by inspection that, when $\tilde r<\tilde\alpha$, the presence of the limit cycle at threshold makes the system periodically flip between \emph{six} possible phase-locked configurations, namely, $(0\pi\pi)$, $(0\pi0)$, $(000)$, $(\pi00)$, $(\pi0\pi)$, and $(\pi\pi\pi)$, corresponding to the six degenerate ground-state configurations of the frustrated Ising model. One of these configurations stabilizes the long-time dynamics \emph{only} when $\tilde r=0$ (see figure~\ref{fig:fixedpointconfigurationsthreeoscillators}, right panel). The frustration of one of the dissipative components of the coupling has therefore a dramatic effect on the coherent dynamics of the network. In the non-frustrated case, the system at threshold converges to a phase-locked configuration for $\tilde r<\tilde\alpha$, and displays persistent beats otherwise. Instead, the behaviour of the fully-frustrated network is \emph{reversed} with respect to the non-frustrated case: It presents beats for $\tilde r<\tilde\alpha$, and converges to phase-locked oscillations otherwise. For fixed $\tilde\alpha$, the frequency of the beats increases linearly from zero for small $\tilde r$, and goes to zero as $\tilde r$ approaches $\tilde\alpha$ with the same critical exponent $1/3$ as before.

\begin{figure}[t]
\centering
\includegraphics[width=11cm]{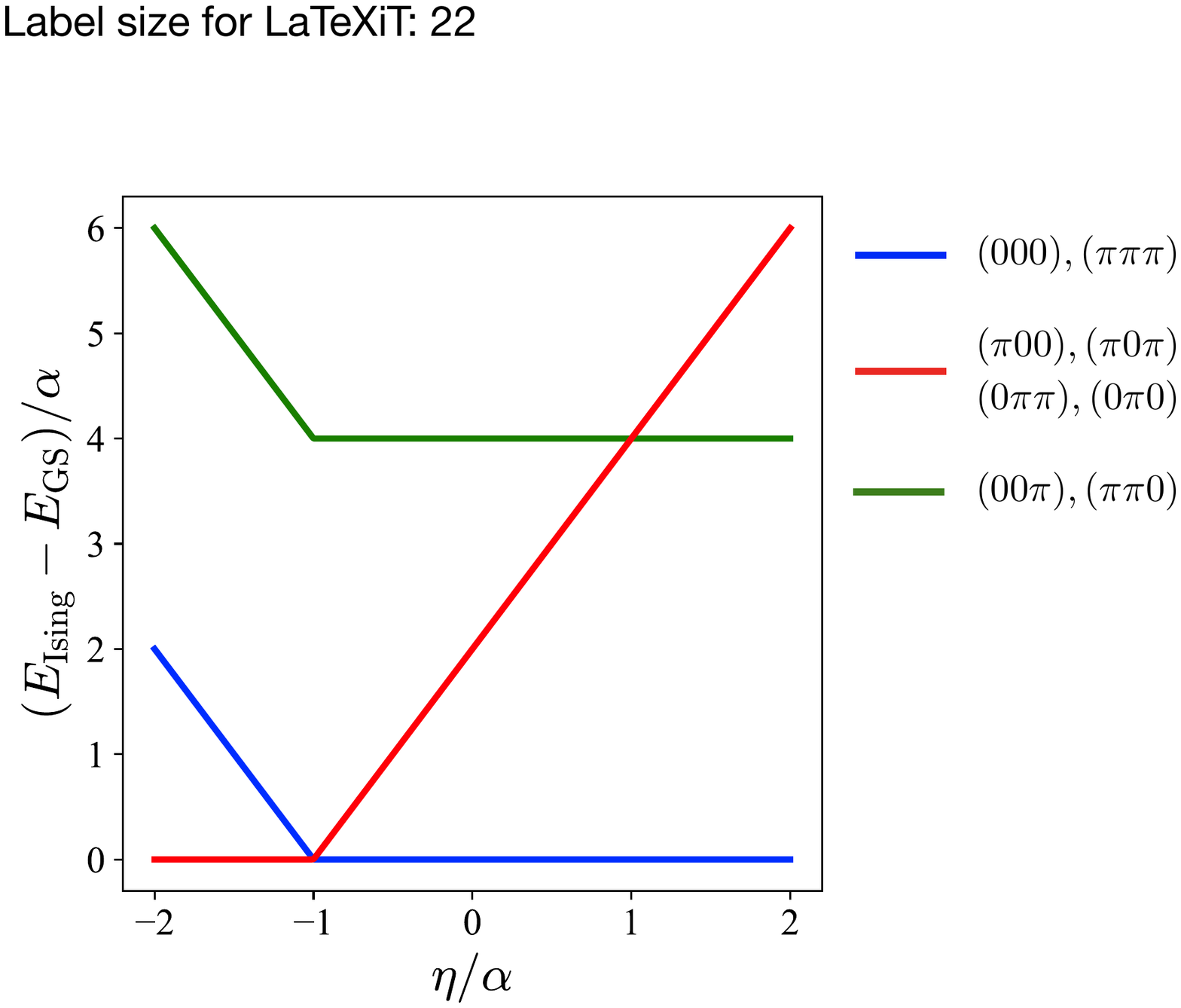}
\caption{Ising energy levels relative to the ground-state energy as a function of $\eta/\alpha$, from the classical Ising energy that the dissipative components of the coupling reflect: $E_{\rm ising}=-\eta\,\sigma_1\sigma_2-\alpha(\sigma_1\sigma_3+\sigma_2\sigma_3)$, where $\sigma_j=\pm1$ is the Ising spin variable corresponding to the phase solution (respectively ``$0$'' or ``$\pi$'') of $x_j$. The three energy levels are shown in different colors for different states: blue for $(000)$ and $(\pi\pi\pi)$, red for $(\pi00)$, $(\pi0\pi)$, $(0\pi\pi)$, and $(0\pi0)$, and green for $(00\pi)$ and $(\pi\pi0)$. The Ising gap goes to zero at the fully-frustrated point $\eta/\alpha=-1$, where the six states $(\pi00)$, $(\pi0\pi)$, $(0\pi\pi)$, $(0\pi0)$, $(000)$ and $(\pi\pi\pi)$ become degenerate.}
\label{fig:isingenergies}
\end{figure}

\subsection{Interpolating case}
\label{sec:generalcase}
We now expand the discussion to consider the general case of equation~\eqref{eq:couplingmatrixthreecoupledparametricoscillators2}, which interpolates between the non-frustrated and the frustrated cases, and generalizes the analysis in sections~\ref{sec:nonfrustratednetwork} and~\ref{sec:fullyfrustratednetwork}. Here, one can find the frequency of the beats at threshold $\omega_B$ from the imaginary part of the most efficient eigenvalue, for a fixed $\tilde\alpha$, as a function of $\tilde r$ and $\tilde\eta$, and discern regions in parameter space \revsecond{where} phase-locked oscillations or beats are observed. We present our findings in figure~\ref{fig:frequencybeatsrvseta} for both theory [panel \textbf{(a)}] and low-level simulation of the experiment [panel \textbf{(b)}], which will be discussed in detail in the next section. To ease the comparison between the analytical prediction and the low-level simulation results, we express the frequency of the beats in units of $\alpha$. This makes the frequency of the beats $\omega_{\rm B}/\alpha$ be a function of $r/\alpha$ multiplied by pure numbers and independent of $g$ (see sections~\ref{sec:nonfrustratednetwork} and~\ref{sec:fullyfrustratednetwork}, and~\ref{appendix:jacobianmatrixspectrumintheinterpolatingcase}).

Panel \textbf{(a)} of figure~\ref{fig:frequencybeatsrvseta} shows $\omega_{\rm B}/\alpha$ in the $r/\alpha$ vs. $\eta/\alpha$ plane, where the red lines mark the special cases of fully frustrated and totally non-frustrated Ising coupling ($\eta/\alpha=\pm1$) that were discussed in sections~\ref{sec:nonfrustratednetwork} and~\ref{sec:fullyfrustratednetwork}. Green boundaries separate the regions where phase-locking is found ($\omega_B=0$, dark blue) from those where persistent beating is manifested ($\omega_B>0$, other colors). We see that, for $\eta/\alpha>0$, a region of beats is found when the energy-preserving coupling dominates over the dissipative coupling ($ r>\alpha$), and phase-locking is found otherwise, similar to the two-oscillator analysis and to the non-frustrated case. However, when $\eta/\alpha<0$, a ``tooth-shaped'' region of beats appears when the dissipative coupling dominates ($ r<\alpha$), and phase locking is found otherwise. As $r$ is lowered towards $ r=0$, the width of the tooth region of beats decreases until it collapses to a point when $ r=0$, at $\eta=-\alpha$, i.e., the fully-frustrated point of section~\ref{sec:fullyfrustratednetwork}.

This peculiar behaviour of beating for small $r/\alpha$ may have important implications in the context of \revsecond{PO-CIM}s. For example, fixing $r$ and scanning $\eta$ from positive to negative allows to interpolate between the ferromagnetic non-frustrated ($\eta = \alpha$), and fully frustrated ($\eta = -\alpha$) Ising models, where the transition to the fully-frustrated case occurs at $\eta/\alpha=-1$. At the transition point, the Ising gap (i.e., the energy difference between the ground- and first-excited configurations) closes, causing the multiplicity of the ground state to increase (in our case, from two to six, see blue and red curves in figure~\ref{fig:isingenergies}). Our findings \revsecond{show} that\revsecond{, in the system of three parametric oscillators coupled by the matrix in Eq.~\eqref{eq:couplingmatrixthreecoupledparametricoscillators2},} any \emph{vanishingly small} energy-preserving coupling induces coherent beating between the oscillators as the Ising gap becomes vanishingly small, preventing the system from converging to the Ising ground-state configuration.

In addition, we find that, while \revsecond{our} three-oscillator \revsecond{system} correctly behaves as a \revsecond{PO-CIM} at the fully-frustrated point and in the phase-locking ``Ising'' region in figure~\ref{fig:frequencybeatsrvseta}, in the other phase-locking (``Non Ising'') region, the system does not yield the expected Ising behaviour. Indeed, the Ising model predicts a \emph{four-fold} degenerate ground-state when $\eta<-\alpha$ (see red curve in figure~\ref{fig:isingenergies}). However, in the ``Non Ising'' region in figure~\ref{fig:frequencybeatsrvseta}, the oscillator system slightly above the threshold converges only to \emph{two} fixed points. In particular, we find the following behavour:
\begin{itemize}
\item For $r=0$ and $r=\alpha$, the two fixed points are found on the $X_{3,R}=0$ plane, implying that $\lim_{t\rightarrow\infty}x_3(t)=0$, and the other two oscillators converge to $(0\pi)$ or $(\pi0)$. Clearly, the suppression of one oscillator in the long-time limit is not a valid Ising configuration;
\item For $0<r<\alpha$, the two fixed points correspond to the states $(0\pi\pi)$ and $(\pi00)$, which are only two of the four ground states of the Ising model;
\item For $r>\alpha$, the two fixed points correspond to the states $(0\pi0)$ and $(\pi0\pi)$. These two configurations, for $\eta<-\alpha$, are two of the four ground states, as before, but for $-\alpha<\eta<0$, they correspond to two of the four \emph{excited} states of the Ising model.
\end{itemize}
This finding hints that frustration may cause phase-locked oscillation at threshold that however transcend the Ising description. \revsecond{Before concluding, we stress that the presence and details of the beating region for vanishingly small energy-preserving coupling strongly depend on the form of the coupling matrix, as well as on the number of oscillators. Indeed, while frustrated spin models with a small number of spins, simulated with PO-CIMs, have been experimentally studied in previous work~\cite{marandi2014cim,takata2016cubic}, coherent beats were not reported. This fact can be due to both the different coupling matrix considered in~\cite{marandi2014cim,takata2016cubic}, as well as to the fact that the energy-preserving coupling was possibly suppressed.} A deeper analysis on \revsecond{how our results translates to general number of coupled oscillators and general coupling topology} requires further analysis of larger spin models, which is beyond the scope of the present manuscript, and it is left for future work.

\section{Numerical simulation of the experimental implementation}
\label{sec:experimentalrealization}
To corroborate the previous results and confirm our analytical predictions, we conducted a direct numerical simulation of the dynamics of the field inside a parametric-oscillator cavity with three (or more) modes, aiming to emulate as closely as possible the dynamics of a future experimental setup. In such an experiment, we intend to couple between parametrically driven modes of a multi-mode radio-frequency cavity in a fully-controlled and tunable manner. The dynamical coupling will be controlled by a field-programmable gate array (FPGA). Full details of the planned actual implementation will be reported in future work. Our numerical approach is completely distinct from the analysis presented above, and makes no explicit mention (or use) of the coupled Matheiu's equations of motion. In what follows, we first discuss our numerical procedure, and then compare the results of the simulated experiment with the analytical results presented in the previous sections.

\subsection{Numerical procedure}
\begin{figure}[t]
	\centering
	\includegraphics[width=14cm]{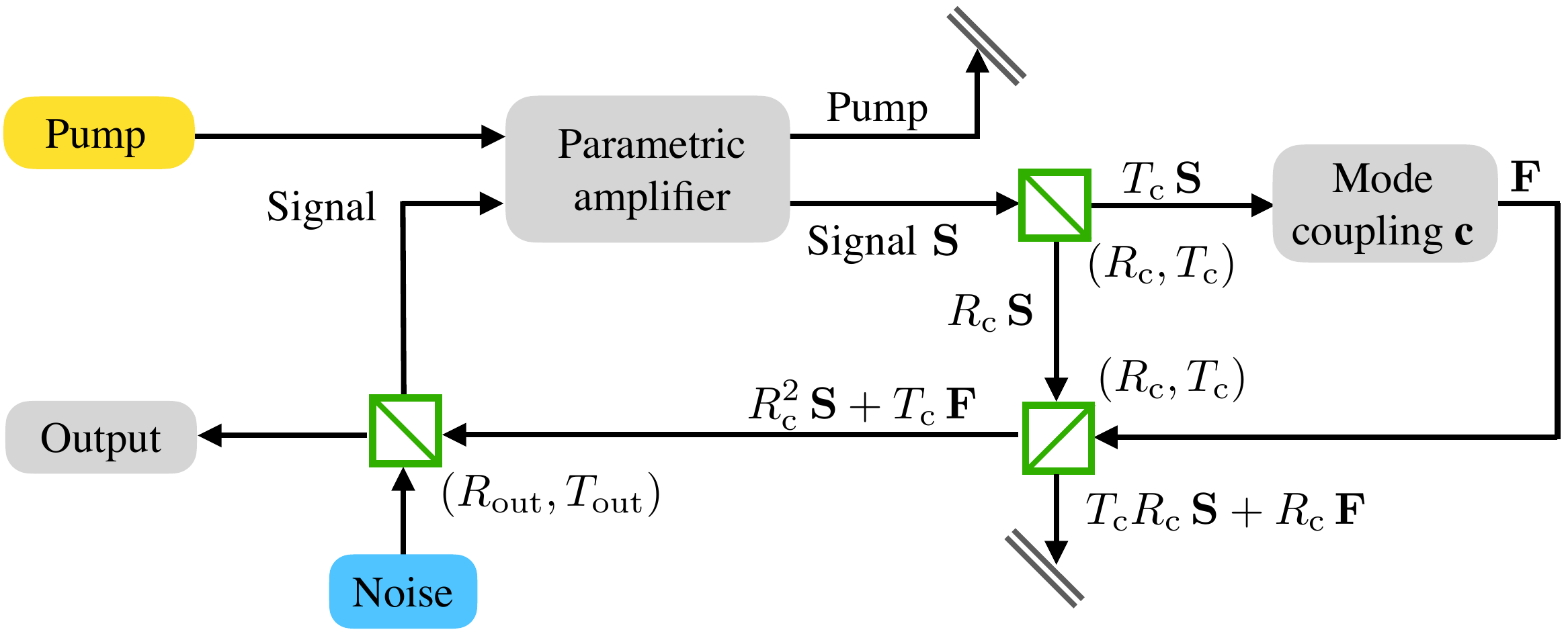}
	\caption{Schematic of the simulated experiment. At each round trip, the field inside the cavity is fed with noise and injected together with the pump field into a parametric amplifier. After the parametric amplification, part of the signal is sent into the coupling mechanism that couples between the time slots, and injected back into the cavity. At the end of the round trip, part of the signal is extracted from the cavity and measured. The green backslashed boxes denote a coupler device, identified by reflection and transmission coefficients $(R,T)$, whose values are chosen differently depending on the simulation step [$(R_{\rm c},T_{\rm c})$ for mode coupling, or $(R_{\rm out},T_{\rm out})$ for output coupling and noise feeding].}
	\label{fig:simulationscheme}
\end{figure}

We consider a multimode cavity where each temporal slot acts as an independent parametric oscillator, dynamically coupled to the other modes. In our simulation, the field inside the cavity propagates as illustrated in the block diagram in figure~\ref{fig:simulationscheme}. At each round trip inside the cavity, the time signal is partitioned into $N=3$ time slots, and parametrized as a three-dimensional vector $\mathbf{S}={(S_1,S_2,S_3)}^T$. In each such interval, the field is assumed to vary slowly and is amplified independently of the other time slots. This is a reasonable assumption since the parametric gain is an instantaneous process and the pump for each time slot is uncoupled from that of the other slots. Thus, each time slot defines a distinct parametrically driven mode that without coupling evolves independently from the others. Furthermore, additive noise is fed inside the cavity at each round trip through an output coupler device to simulate thermal noise or vacuum fluctuations. During the first round trip, before being injected into the parametric amplifier, the signal inside the cavity consists of noise alone.

A round trip inside the cavity is identified by the following steps: First, the pump field and the signal are injected into the parametric amplifier. Importantly, since our goal is to probe the linear, near threshold, properties of the system, the pump intensity is set slightly above the oscillation threshold (section~\ref{sec:threecoupledparametricoscillators}). At the output of the parametric amplifier, the residual pump field is blocked (dark grey parallel lines in figure~\ref{fig:simulationscheme}) and the signal is injected into a coupler, which splits the field according to transmission and reflection coefficients $T_{\rm c}=1/4$ and $R_{\rm c}=\sqrt{1-T_{\rm c}^2}$. The transmitted signal $T_{\rm c}\mathbf{S}$ is sent into a coupling mechanism, which implements the coupling matrix $\mathbf{c}$ of equation~\eqref{eq:couplingmatrixthreecoupledparametricoscillators2}. In a future experiment, such a coupling mechanism will be implemented by an FPGA. After the coupling, the coupled signal $\mathbf{F}=T_{\rm c}\,\mathbf{c}\,\mathbf{S}$ and the reflected signal $R_{\rm c}\mathbf{S}$ are combined on another coupling device, again with transmission and reflection coefficients $T_{\rm c}$ and $R_{\rm c}$. At this step, the reflected signal $T_{\rm c}R_{\rm c}\mathbf{S}+R_{\rm c}\mathbf{F}$ is blocked (contributing to the overall cavity losses), and the transmitted signal $R_{\rm c}^2\mathbf{S}+T_{\rm c}\mathbf{F}$ is fed back into the cavity. Last, the output coupler with transmission and reflection coefficients $T_{\rm out}=1/2$ and $R_{\rm out}=\sqrt{1-T_{\rm out}^2}$ allows to couple out the field $T_{\rm out}(R_{\rm c}^2\mathbf{S}+T_{\rm c}\mathbf{F})$ from the cavity and analyze it both it in time and frequency, and the remaining field $R_{\rm out}(R_{\rm c}^2\mathbf{S}+T_{\rm c}\mathbf{F})$, fed with noise, is input together with the pump field into the parametric amplifier to start the next round trip.

For a given set of parameters, we run the simulation over a sufficient number of round trips $n_{\rm rt}$  in order for the oscillation to reach a steady state. We then examine the steady-state dynamics to obtain the slow-varying amplitude of all oscillators and numerically extract the frequency of the beats.

\subsection{Numerical results}
\label{sec:numericalresults}
The numerical frequency of the beats is measured by computing the fast Fourier transform (FFT) of the signal at the output to identify the frequency component with largest amplitude, for different values of $r/\alpha$ and/or $\eta/\alpha$. In the numerics, the frequencies obtained from the FFT are given in units of $1/\tau_{\rm sim}$, where the total simulation time is $\tau_{\rm sim}=n_{\rm rt}\tau_{\rm rt}$, and round-trip time $\tau_{\rm rt}$ in our numerical context is an arbitrary time scale. In order to quantitatively compare the numerical results with the analytical prediction, we express the frequency of the beats in units of $\alpha$ (section~\ref{sec:generalcase}) and use $\tau_{\rm rt}=0.0195$ (see \ref{appendix:detailsonthefit} for more details).

\begin{figure}[t]
\centering
\includegraphics[width=15.7cm]{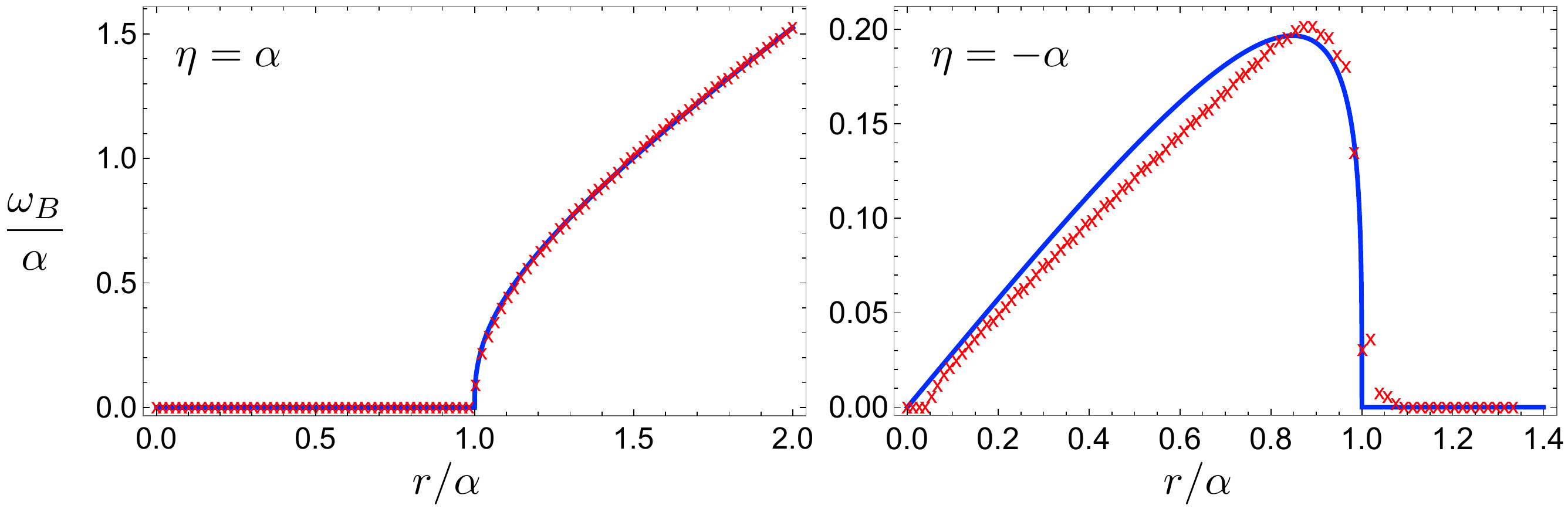}
\caption{Frequency of the beats at threshold $\omega_{\rm B}/\alpha$ as a function of $r/\alpha$, (Left) in the non-frustrated network ($\eta=\alpha$), and (Right) in the fully-frustrated network ($\eta=-\alpha$). The numerical data (red points) obtained from the simulation of the experiment with $\tau_{\rm rt}=0.0195$ are superimposed to the analytical curve [blue line, equations~\eqref{eq:frequencyofthebeatsnonfrustrated} and~\eqref{eq:frequencyofthebeatsfrustrated}].}
\label{fig:frequencyofthebeatsnumericstheory}
\end{figure}

Figure~\ref{fig:frequencyofthebeatsnumericstheory} shows the frequency of the beats as a function of $r/\alpha$ for both the numerical simulation of the experiment and the theoretical analysis. We compare the results in the special cases of the non-frustrated network ($\eta=\alpha$) and the fully-frustrated one $(\eta=-\alpha)$. As evident, the numerical results in both cases agree well with the theoretical curves, with some slight deviations especially in the fully-frustrated cases, which we ascribe to the difficulty of correctly estimating the oscillation threshold and therefore choosing the proper value of $h$, due to both noise and nonlinearities. Indeed, it has been shown that pump-depletion nonlinearity tends to lower the beating frequency (divergence of the period of the beats), eventually inducing phase-locking as $h$ is increased above the oscillation threshold~\cite{PhysRevLett.123.083901}.

Last, in figure~\ref{fig:frequencybeatsrvseta}, panel \textbf{(b)}, we evaluate the beating frequency $\omega_{\rm B}/\alpha$ as a function of $r/\alpha$ and $\eta/\alpha$, in order to verify the theoretical phase diagram in panel \textbf{(a)}. The analytical phase boundary, marked by the green line, which is the same for both panels of figure~\ref{fig:frequencybeatsrvseta}, is superimposed to the numerical phase diagram to ease comparison between theory and simulated experiment. As evident, our numerical data agree exceptionally well with the analytical prediction also in the interpolating case.


\section{Conclusions}
\label{sec:conclusions}
We analyzed the behaviour of three coupled degenerate parametric oscillators - the minimal case to study nontrivial coupling and connectivity effects. By extending our previous work on two coupled parametric oscillators, we modelled the system as three coupled Mathieu's equations, where the coupling between any two oscillators is comprised of both energy-preserving and dissipative components. \revfirst{We analyzed the role of frustration of the dissipative component of the coupling for specific choices of the coupling matrix}. We focused in particular on two main cases of connectivity, namely, for frustrated and non-frustrated dissipative coupling. Our theoretical predictions, obtained by linearizing the effective equations of motion, were confirmed by a direct numerical simulation in time of the dynamics inside a parametric oscillator cavity, as it would be implemented in an actual experiment. The good agreement between the results obtained by these two different approaches strengthens the fact that the coupled nonlinear Mathieu's equations capture the relevant dynamics of coupled parametric oscillators.

Our main finding was that frustration of the dissipative component of the coupling has a dramatic effect on the coherent dynamics of the system. While in the non-frustrated case the system phase locks once the dissipative coupling exceeds the energy-preserving one, and behaves as a \revsecond{PO-CIM}, in qualitative agreement with the behaviour of two coupled parametric oscillators, the frustrated case shows a totally reversed behaviour. In particular, when the dissipative coupling dominates close to full frustration, the Ising gap is vanishingly small and any vanishingly small energy-preserving coupling induces coherent beats between the (quasi) degenerate Ising configurations. For large values of the frustration parameter and for large energy-preserving coupling, the system phase locks. Interestingly, in this phase-locking region the system of three coupled oscillators does not obey the Ising description.

Our results provide an additional piece of evidence of the highly nontrivial dynamics in networks of coupled parametric oscillators, which is considerably richer than an Ising network of spins. In the view of using coupled parametric oscillators to simulate Ising models, our results hint that, in situations where the energy gap of the corresponding Ising model is very small or vanishes, the presence of even a small energy-preserving coupling between the oscillators \revsecond{may} induce coherent beats, \revsecond{which should be considered in the context of PO-CIMs}. Because of these intriguing implications, the theoretical and experimental investigation of large-scale networks is now highly desirable in order to \revsecond{see how} our results \revsecond{translate to} larger sets of parametric oscillators, \revfirst{as well as for more general forms of the coupling matrix~\cite{hamerlyfristratedchain2016}}. We are currently planning the experimental implementation of such a large-scale network, whose analysis will be reported in future work. Ultimately, in light of our findings and the correspondence between the phase-locked behavior and classical time-crystals, it will be important to understand how nontrivial connectivities affect the stability of classical many-body time crystals.


\section*{Acknowledgements}
We thank Itzhack Dana for fruitful discussions. A. P. acknowledges support from the Israel Science Foundation (ISF) Grants No. 44/14 and U.S.-Israel Binational Science Foundation (BSF) Grant No. 2017743.  M.~C.~S. acknowledges support from the ISF Grants No.~231/14, 1452/14, and~993/19, and BSF Grants No.~2016130 and~2018726.


\appendix

\section{Derivation of the power-splitter coupling}
\label{appendix:derivationofthepowersplittercoupling}
In this appendix, we explicit the origin of the power-splitter coupling as in equation~\eqref{eq:powersplittercouplingmatrix}. In an actual physical implementation, the parametric oscillators are realized by a nonlinear cavity. The fields $x_1$ and $x_2$ inside each cavity propagate with a characteristic round-trip time $\tau_{\rm rt}=D/v$, which depends on the linear dimension $D$ of the cavity, and on the field propagation velocity $v$ inside the cavity.

The fields after $n+1$ round-trip times, $t_{n+1}$ relate to the fields after $n$ round-trip times, $t_n=n\,\tau_{\rm rt}$, for $n=0,1,2,\ldots$, via the splitter matrix as
\begin{equation}
\left(\begin{array}{c}x_1(t_{n+1})\\x_2(t_{n+1})\end{array}\right)=\left(\begin{array}{cc}c_{11}&c_{12}\\-c_{21}&c_{22}\end{array}\right)\left(\begin{array}{c}x_1(t_{n})\\x_2(t_{n})\end{array}\right) \,\, .
\label{eq:equationforthefieldsinsidethecavities111}
\end{equation}
We consider $c_{11}=c_{22}\equiv c$, where $0\leq c\leq 1$ represents the transmittance coefficient of the coupling. With this choice, equation~\eqref{eq:equationforthefieldsinsidethecavities111} is equivalently recast as
\begin{equation}\left\{
\begin{array}{l}x_1(t_{n+1})=c\,x_1(t_n)+c_{12}\,x_2(t_n)\\\\
x_2(t_{n+1})=c\,x_2(t_n)-c_{21}\,x_1(t_n)\end{array} \,\, ,
\right.
\label{eq:equationforthefieldsinsidethecavities11112132}
\end{equation}
and by rewriting $c\,x_{1,2}=x_{1,2}-(1-c)x_{1,2}$, equation~\eqref{eq:equationforthefieldsinsidethecavities11112132} becomes
\begin{equation}\left\{
\begin{array}{l}x_1(t_{n+1})=x_1(t_n)-(1-c)\,x_1(t_n)+c_{12}\,x_2(t_n)\\\\x_2(t_{n+1})=x_2(t_n)-(1-c)\,x_2(t_n)-c_{21}\,x_1(t_n)\end{array}\right. \,\, .
\label{eq:equationforthefieldsinsidethecavities11}
\end{equation}
Since $x_{1,2}(t_{n+1})-x_{1,2}(t_{n})\propto\dot x_{1,2}(t)/\omega_0$, equation~\eqref{eq:equationforthefieldsinsidethecavities11} can be rewritten as
\begin{equation}\left\{
\begin{array}{l}\dot x_1=-\omega_0\,(1-c)\,x_1+\omega_0\,c_{12}\,x_2\\\\\dot x_2=-\omega_0\,(1-c)\,x_2-\omega_0\,c_{21}\,x_1\end{array}\right. \,\, .
\label{eq:equationforthefieldsinsidethecavities211}
\end{equation}
Without loss of generality, we consider $c_{12}>0$. The terms proportional to $1-c$ in equation~\eqref{eq:equationforthefieldsinsidethecavities211} can be seen as loss terms that can be absorbed into the definition of $g$, the intrinsic loss of the cavities. Therefore, by taking the time derivative on both sides of equation~\eqref{eq:equationforthefieldsinsidethecavities211}, and by including this coupling in the equations of motion, equation~\eqref{eq:equationsofmotionfortwooscillators} is obtained.

\section{Hamiltonian for the power-splitter coupling}
\label{appendix:hamiltonianforthepowersplittercoupling}
In this appendix, we report the derivation of the equations of motion~\eqref{eq:equationsofmotionfortwooscillators} (with $\beta=0$ and $g=0$) from the Hamilton's equations, in order to show that the coupling with $c_{12}=c_{21}=r$ is indeed energy preserving. First, one takes the two oscillators fields, $x_1$ and $x_2$, and their conjugate momentum variables, $p_1$ and $p_2$, and defines the vectors of canonical coordinates $\mathbf{p}={(p_1,p_2)}^T$ and $\mathbf{x}={(x_1,x_2)}^T$, where $T$ denotes the transposition. The Hamiltonian of the system is analogous to that of a particle of charge $q$ in a two-dimensional plane, in a vector potential along the $z$-axis (i.e., perpendicular to the plane), given by $\mathbf{A}={(-x_2,x_1,0)}^T$, so that the corresponding effective magnetic field is $\mathbf{B}=\nabla\times\mathbf{A}=\hat z\left(\partial_{x_1}A_{x_2}-\partial_{x_2}A_{x_1}\right)=\hat z\,2$:
\begin{eqnarray}
H&=&\frac{1}{2m}{\left(\mathbf{p}-q\,\mathbf{A}\right)}^2+\frac{1}{2}\,m\omega^2_0\left[1+h\,\sin(2\omega_0t)\right]\mathbf{x}^2\nonumber\\
&=&\frac{\mathbf{p}^2}{2m}+\frac{1}{2}\,m\omega^2_0[1+h\,\sin(2\omega_0t)\mathbf{x}^2-\frac{q}{m}\,\mathbf{p}\cdot\mathbf{A}+\frac{q^2}{2m}\,\mathbf{A}^2 \,\, ,
\end{eqnarray}
or explicitly in terms of the canonical variables $x_1$, $x_2$, and $p_1$, $p_2$
\begin{equation}
\hspace{-1.5cm}H=\frac{p^2_1+p^2_2}{2m}+\frac{1}{2}\,m\,\omega^2_0[1+h\,\sin(2\omega_0t)](x_1^2+x_2^2)+\frac{q^2}{2m}\,(x^2_1+x_2^2)+\frac{q}{m}(p_1x_2-p_2x_1) \,\, .
\end{equation}
The Hamilton's equations~\cite{landau1982mechanics} for the $\{\dot x\}$ variables are
\begin{equation}
\dot x_1=\frac{\partial H}{\partial p_1}=\frac{p_1}{m}+\frac{q}{m}\,x_2 \qquad \dot x_2=\frac{\partial H}{\partial p_2}=\frac{p_2}{m}-\frac{q}{m}\,x_1  \,\, ,
\label{eq:equationsofmotionforxhamiltonequations}
\end{equation}
and the Hamilton's equations for the $\{\dot p\}$ variables are
\begin{eqnarray}
\dot p_1=-\frac{\partial H}{\partial x_1}=-m\omega^2_0[1+h\,\sin(2\omega_0t)]x_1-\frac{q^2}{m}\,x_1+\frac{q}{m}\,p_2\nonumber\\
\dot p_2=-\frac{\partial H}{\partial x_2}=-m\omega^2_0[1+h\,\sin(2\omega_0t)]x_2-\frac{q^2}{m}\,x_2-\frac{q}{m}\,p_1 \,\, .
\label{eq:equationsofmotionforphamiltonequations}
\end{eqnarray}
From equation~\eqref{eq:equationsofmotionforxhamiltonequations}, by deriving both sides with respect to time, one has
\begin{equation}
\dot p_1=m\,\ddot x_1-q\,\dot x_2 \qquad \dot p_2=m\,\ddot x_2+q\,\dot x_1 \,\, .
\label{eq:equationsofmotionforxhamiltonequations2}
\end{equation}
By substituting $\dot p_1$ and $\dot p_2$ in the left-hand sides of equation~\eqref{eq:equationsofmotionforxhamiltonequations2} with the expressions in equation~\eqref{eq:equationsofmotionforphamiltonequations}, one has
\begin{eqnarray}
m\,\ddot x_1-q\,\dot x_2=-m\omega^2_0[1+h\,\sin(2\omega_0t)]x_1-\frac{q^2}{m}\,x_1+\frac{q}{m}\,p_2\nonumber\\
m\,\ddot x_2+q\,\dot x_1=-m\omega^2_0[1+h\,\sin(2\omega_0t)]x_2-\frac{q^2}{m}\,x_2-\frac{q}{m}\,p_1 \,\, ,
\label{eq:equationsofmotionforxhamiltonequations3}
\end{eqnarray}
but, from equation~\eqref{eq:equationsofmotionforxhamiltonequations}, one has $p_1/m=\dot x_1-(q/m)x_2$ and $p_2/m=\dot x_2+(q/m)x_1$ that, when substituted in the right-hand side of equation~\eqref{eq:equationsofmotionforxhamiltonequations3}, yields
\begin{equation}
\begin{array}{l}
m\,\ddot x_1-q\,\dot x_2=-m\omega^2_0[1+h\,\sin(2\omega_0t)]x_1+q\,\dot x_2\\\\
m\,\ddot x_2+q\,\dot x_1=-m\omega^2_0[1+h\,\sin(2\omega_0t)]x_2-q\,\dot x_1 \,\, ,
\end{array}
\label{eq:equationsofmotionforxhamiltonequations4}
\end{equation}
from which one obtains the equations of motion
\begin{equation}
\begin{array}{l}
\ddot x_1+\omega^2_0[1+h\,\sin(2\omega_0t)]x_1-(2q/m)\,\dot x_2=0\\\\
\ddot x_2+\omega^2_0[1+h\,\sin(2\omega_0t)]x_2+(2q/m)\,\dot x_1=0 \,\, ,
\end{array}
\label{eq:equationofmotionmodel1vectorpotential}
\end{equation}
which are indeed the equations of motion in equation~\eqref{eq:equationsofmotionfortwooscillators} with $c_{12}=c_{21}=r$, $\beta=0$, and $g=0$, where $\omega_0r=2q/m$.

\section{Jacobian matrix spectrum in the interpolating case}
\label{appendix:jacobianmatrixspectrumintheinterpolatingcase}
In this appendix, we report the expressions of the eigenvalues of the Jacobian matrix around the origin in the interpolating case discussed in section~\ref{sec:generalcase} (see also figure~\ref{fig:frequencybeatsrvseta}). By generalizing the functions in equation~\eqref{eq:expressionsforfandg}, one has
\begin{equation}
\hspace{-1cm}G(\tilde r,\tilde\alpha,\tilde\eta)=\frac{1}{2}\,{\left[\tilde \eta\left(\tilde r^2- \tilde \alpha ^2  \right) +\sqrt{\left(\tilde r^2-\cfrac{2 \tilde \alpha ^2+ \tilde \eta ^2}{3}\right)^3+\eta^2\left(\tilde r^2- \tilde \alpha ^2  \right)^2}\,\right]}^{1/3}
\end{equation}
\begin{equation}
\hspace{-1cm}F(\tilde r,\tilde\alpha,\tilde\eta)=\frac{\tilde r^2-\left(2 \tilde \alpha ^2+ \tilde \eta ^2\right)/3}{4\,G(\tilde r,\tilde\alpha,\tilde\eta)} \,\, ,
\end{equation}
and then the eigenvalues of the Jacobian that can have a positive real part can be written as
\begin{equation}
\lambda_{1}(\tilde r,\tilde\alpha,\tilde\eta)=-\frac{1}{2}+\frac{\tilde h}{4}+ F(\tilde r,\tilde\alpha,\tilde\eta)- G(\tilde r,\tilde\alpha,\tilde\eta) \label{eq:lambda1pmthreeparametricoscillators1212}
\end{equation}
\begin{equation}
\lambda_{2}(\tilde r,\tilde\alpha,\tilde\eta)=-\frac{1}{2}+\frac{\tilde h}{4}-e^{i\pi/3}\,F(\tilde r,\tilde\alpha,\tilde\eta)+e^{-i\pi/3}\,G(\tilde r,\tilde\alpha,\tilde\eta)\label{eq:lambda2pmthreeparametricoscillators1212}
\end{equation}
\begin{equation}
\lambda_{3}(\tilde r,\tilde\alpha,\tilde\eta)=-\frac{1}{2}+\frac{\tilde h}{4}-e^{-i\pi/3}\,F(\tilde r,\tilde\alpha,\tilde\eta)+e^{i\pi/3}\,G(\tilde r,\tilde\alpha,\tilde\eta) \,\, .\label{eq:lambda3pmthreeparametricoscillators1212}
\end{equation}
By finding the most efficient eigenvalue $\lambda_{\rm max}(\tilde r,\tilde\alpha,\tilde\eta)$ according to the usual condition (the eigenvalue with largest real part), the frequency of the beats at threshold reads $\omega_{\rm B}(\tilde r,\tilde\alpha,\tilde\eta)=\omega_0g\,|{\rm Im}[\lambda_{\rm max}(\tilde r,\tilde\alpha,\tilde\eta)]|$.

\section{Additional details on the choice of the round-trip time}
\label{appendix:detailsonthefit}
\begin{figure}[t]
\centering
\includegraphics[width=15.7cm]{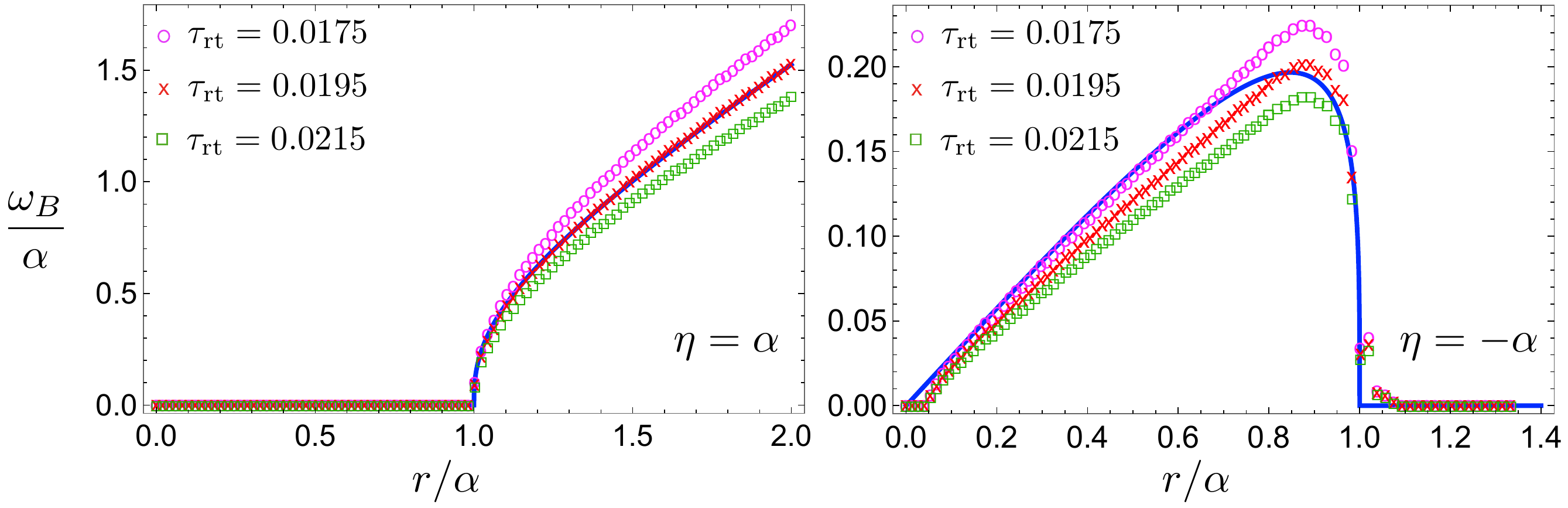}
\caption{Beat frequency at threshold $\omega_{\rm B}$ as a function of $r/\alpha$, (Left panel) in the non-frustrated case and (Right panel) fully-frustrated case, as in figure~\ref{fig:frequencyofthebeatsnumericstheory}. The data from the simulated experiment for three different values of $\tau_{\rm rt}$ as in the legends (purple circles for $\tau_{\rm rt}=0.0175$, red crosses for $\tau_{\rm rt}=0.195$, and green squares for $\tau_{\rm rt}=0.0215$) are compared to the theoretical behaviour [blue line, equations~\eqref{eq:frequencyofthebeatsnonfrustrated} and~\eqref{eq:frequencyofthebeatsfrustrated}].}
\label{fig:figurecomparisontheorynumerics}
\end{figure}
In this appendix, we provide some details on the choice of $\tau_{\rm rt}$ discussed in figure~\ref{fig:frequencyofthebeatsnumericstheory}. In figure~\ref{fig:figurecomparisontheorynumerics}, we show the comparison between the theoretical expressions of the frequency of the beats at threshold [equations~\eqref{eq:frequencyofthebeatsnonfrustrated} and~\eqref{eq:frequencyofthebeatsfrustrated}] $\omega_{\rm B}/\alpha$, as a function of $r/\alpha$, and the data from the simulated experiment for different values of the round-trip time $\tau_{\rm rt}$, as in the legends. The effect of changing $\tau_{\rm rt}$ is to renormalize the frequency units for the simulated experiment. Because of the excellent agreement between theory and data from the simulated experiment in the non-frustrated case for $\tau_{\rm rt}=0.0195$, this value of $\tau_{\rm rt}$ was chosen to quantitatively match the theoretical phase diagram in figure~\ref{fig:frequencybeatsrvseta}. Indeed, as evident, for this $\tau_{\rm rt}$, theory and data are essentially overlapped.


\section*{References}
\providecommand{\newblock}{}

\end{document}